# Top-Down Causation by Information Control:
# From a Philosophical Problem to a Scientific Research Program

## G. Auletta[*], G. F. R. Ellis[+], FRS, and L. Jaeger[°]

**Abstract.** *It has been claimed that different types of causes must be considered in biological systems, including top-down as well as same-level and bottom-up causation, thus enabling the top levels to be causally efficacious in their own right. To clarify this issue, important distinctions between information and signs are introduced here and the concepts of information control and functional equivalence classes in those systems are rigorously defined and used to characterise when top down causation by feedback control happens, in a way that is testable. The causally significant elements we consider are equivalence classes of lower level processes, realised in biological systems through different operations having the same outcome within the context of information control and networks.*

## 1. Introduction

At the most general level, the issue of top-down causation can be seen as the problem of how higher levels of reality (roughly, levels of greater complexity[1]) can possibly have any causal effectiveness on lower levels [Simon 1962, Campbell 1974]. If it is assumed (according to the standard view in physics) that purely physical effects determine what happens at the lower levels and thereby also completely determine what happens at the higher levels, how can there be any real meaning to higher level causes and effects?

Many scientists consider `top-down causation' not to be real: they believe it is just a complicated way of describing things that in the end confuses the real causal patterns, which are believed to be bottom-up only (see Fig. 1a). It is also assumed that phenomena that are not easily understandable in a bottom-up way today, will be so understood in the future. This approach has been extended to all natural systems thanks to the huge success of the application of reductionist methodology in physics and, in recent decades, in molecular biology and neuroscience. As in Francis Crick's famous dictum: "You, your joys and your sorrows, your memories and your ambitions, your sense of personal identity and free will, are in fact no more than the behavior of a vast assembly of nerve cells and their associated molecules" [Crick 1994]. The emphasis in the phrase "*no more than*" is a denial of the reality of anything additional to the pure assembly of cells and is therefore also a rejection of top-down causation.

---

[*] Pontifical Gregorian University, Rome.
[+] Mathematics Department, University of Cape Town.
[°] University of California, Santa Barbara.

[1] Leaving aside its formal definition as that goes further than the scope of this paper.

A similar point of view comes from some emergence theorists, who suggest that, since complex systems can self-assemble from the bottom up, in biological systems there is no need for influence of the whole on the parts [Holland 1997]. In both cases the suggestion is that the introduction of top-down causation is misleading; the real causal powers reside in the lower-level physics and associated classical chemistry; they alone determine what happens.

Nonetheless, there is a wide literature on the emergence of autonomous higher levels of complexity and the role of top-down causation in the hierarchy of complexity (see Clayton and Davies [2006], Murphy and Stoeger [2007] and references therein), expressing a need felt by many scholars to overcome traditional reductionism. Unfortunately, this discussion is mostly confined to philosophical considerations that has little changed the way scientists (especially physicists and molecular biologists) consider their own work.

In this paper, we try to refine relevant concepts in order to translate a philosophical examination of top-down causation into a scientific program able to make predictions and experimental tests. As we shall see, there are already experiments that go in this direction, though they are not interpreted in the way we articulate here. The inquiry regarding top-down causation is somewhat different if we consider cases where consciousness and intelligence are involved, as distinct from those where they are not included. Additionally, important distinctions occur between the cases where life is involved, and those when only physical and chemical interactions occur. This paper will focus on the case of life at its most elementary level, but not on issues raised by intelligence and intention. We think that the concepts presented herein could be crucial for systematically addressing the problem of emergence of complexity and related evolutionary aspects, which we will consider elsewhere.

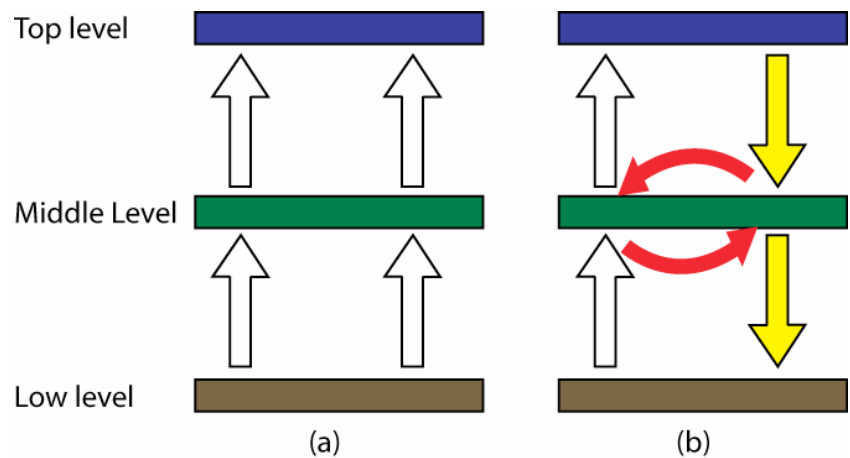

**Figure 1: Bottom-up, Top-down and Same-level modes of action.**
**1(a)** *The classical-standard view understands everything in the universe as happening in terms of bottom-up action only, so that efficient causation is seen essentially as bottom-up.*
**1(b)** *A more careful view is to consider the bottom-up mode of action as providing a space of possibilities, the same-level mode (red arrows) as the true dynamic causation, and the top-*

*down mode (yellow arrows) as changing the causal relations below. Dynamic causation (same-level) can be both efficient and circular. Here, the circular one is shown, since it plays a more crucial role in the context discussed here. While not represented in (a), circular causation can also be part of "bottom-up action only" models.*

---

## 2. A Recent Experiment

A European team [Wegscheid *et al.* 2006] performed *in vivo* experiments of complementation on the bacteria *E. coli* and *B. subtilis*. These bacteria have distinct RNase P enzymes: type A for *E. coli* and type B for *B. subtilis*, respectively. Interestingly, these RNases P have significantly different three-dimensional architectures (see fig.8) that are associated with important biophysical and biochemical differences *in vitro*: for instance, (1) *B. subtilis* P RNA 5' and 3' ends are autolytically processed after association with its protein subunit, while this process is enzyme dependent in *E. coli*; (2)Rnase P from *B. subtilis* but not *E. coli* forms dimers consisting of two RNA and two protein subunits; (3) *B. subtilis* RNase P holoenzyme binds to pre-tRNA with a much higher affinity than mature tRNA, whereas this difference in affinity is attenuated in *E. coli*; (4) type A and type B RNase P RNAs have distinct metal ion requirements.

Now, *despite these structural and functional differences in vitro*, it was shown that types A and B can replace each other *in vivo* without loss in functionality, at least under standard growth conditions. It is interesting to stress that the lower physical-chemical stability of hybrid holoenzyme complexes does not raises functionality problems. Already this shows that the traditional, reductionist point of view is insufficient. Indeed, the concept of function is now widely acknowledged in biology [see also Collins *et al.* 2007].

The authors of the experiment interpret this situation by making the hypothesis that there could be "conserved features" of bacterial RNase P RNA and protein subunits essential for their primary functional activity *in vivo*. While this fact is difficult to deny based on the reality of a consensual core structure for Rnase P enzymes in all living organisms [Altman, 2007; Kirsebom, 2007], we however think that this explanation is only partial and hides a greater truth. In our opinion, their results can be seen in another light, by considering top-down effects through the information control that the unicellular organism exercises on and through these functions. In the following we propose this new conceptual framework.

## 3. What is Top-Down Causation by Information Control?

Generically, top-down causation works by higher-level conditions setting the context for lower level processes (see Bishop and Atmanspacher [2006], Campbell [1974], Clayton and Davies [2006], Murphy and Stoeger [2007], Ellis [2006a, 2006b], Deacon [2006]). Probably, already in physics and chemistry there are boundary conditions that act on lower-level processes. For this reason, top-down causation *applies whatever level is chosen as the reference level*, for there is no known ontologically preferred level (as we do not even know what the lowest level is, there is no known fundamental level). However, in the cases of physics and biology mentioned above, it is not always clear whether the traditional reductionist point of view is actually overcome, since these boundary conditions can again be understood as a complicated effect of much more elementary processes.

Therefore, in this paper, we focus on a specific and stronger kind of top-down causation that plays a very relevant role in biology: top-down causation *by information control*. This does not mean that we think this is the only way top-down causation may be relevant within the biological domain (some other possibilities are discussed in Ellis [2008]). We shall show that top-down causation by information control may be clearly found in biological organisms already at the most elementary *bio*-molecular level, suggesting that there is not the possibility to reinterpret these results in reductionist terms.

**3.1 Some Basic Definitions**
**Top-down causation by information control** is the way *a higher-level instance exercises control of lower level causal processes through feedback control loops making use of functional equivalence classes of operations*. At the biological level, this consists in applying *controlling* signals from the higher level to influence the lower level items' proper mode of action (Fig. 1b). Here, a *signal* is any variation or pattern in a physical or chemical medium that can convey information or be treated as a sign. If we succeed in showing meaningful evidence that this stronger form of top-down causation already happens at the bio-molecular level, we have also succeeded in showing that this occurs in all living systems. Even stronger versions of top-down causation can be found when intentionality and free will are involved; however, as already pointed out, we confine our examination here to the most basic biological domain.

A key concept here is that of *equivalence class of lower-level operations*,[2] discussed in Sec.4 below, where operations occurring in biological systems can be considered as coordinated space-time pathways of physical-chemical interactions. The criterion for an equivalence class of operations is the *outcome* that an operation brings about relative to an established goal: if two different operations give the same outcome, they can be considered equivalent. Thus, what is of concern here are *functional* equivalence classes (sets of operations that produce the same outcome).

Another central point is, of course, information control (Sec.5). In general, information control is the ability to use signals to attain or maintain a specific goal. When different forms of information control are possible, functional equivalence classes are what really counts, since they gather together the possible different operations by which the goal can be attained. Let us stress that any control mechanism is ultimately control in terms of information, even though it makes use of some energy or material to convey that information. Information is consequently a dominant factor for life [Küppers, 1990; Rasmussen et al, 2004; Roederer 2005]. Our point is that the demonstration of the existence in biological systems of functional equivalence classes under information control from above can be seen as strong evidence of top-down causation affecting its lower level modes of operations from a higher level of functional organization.

---

[2] This notion was already introduced into psychology by Lashley [1942] and into neural sciences by Hebb [1949, pp. 38-59], and see also Pearl (1998, 2000).

## 3.2 On the Nature of Causes

The general nature of causation is a key issue, in considering these topics. The majority of scientists think that top-down causation means to act directly on a lower ontological or less complex level of reality without the intermediary of causes acting at this same level. Indeed, a majority of both top-down-causation supporters and detractors assume that it substitutes for the specific modes of action at lower ontological levels. It is easy to see that in this way, the principle of the **closure of the physical world** from the point of view of action and interaction would be violated in most cases, and at least in those where actions according to goals are involved.

To this aim, considering the **nature of causes** and especially distinguishing between *dynamic causes* and *non-dynamic causes* [Auletta 2007d] is of primary importance. We may say that all causes have causal power, but only dynamic causes have causal effectiveness. *Effectiveness* means that the causal agent, in ideal condition, can positively give rise to a certain effect through interactions at the same ontological level, as when a ball moving with a certain speed is able, after collision, to set another ball in motion. *Power* means here concurring in the causal production of something, provided that there also is an effective causal factor. We have effective causal processes when there is an exchange of physical magnitudes. This can be understood, in general, through the so called "transference theory of causation" [Fair 1979, Salmon 1984, Dowe 1992, Salmon 1994, Dowe 1995]

In our language, in fact, **dynamic causes** are same-level *efficient causes*, like thermal energy determining the melting of ice, or *circular causes*, which are those present in non-linear, self-increasing processes, like those commonly occurring in autocatalytic chemical reactions, for instance when a chemical *X* is involved in its own production as in

$$A + X \underset{k_{-1}}{\overset{k_1}{\rightleftharpoons}} 2X,$$

where $k_1$ and $k_{-1}$ are reaction rate constants.

**Non-dynamic causes**, on the contrary, are essentially of two types: *Material causes* (from below), which represent the support for processes and entities at a higher level of complexity, and can therefore be considered as possibility conditions for those processes and entities (for instance the various chemicals constituting the bio-molecules and chemical processes underlying biological processes); and *formal causes* (from above), which are restrictions of the space of possibilities (for instance, during DNA replication, only certain chemical reactions are able to occur because of the context).

**Top-down causes** are causes that *do not act* at the same ontological level (see Fig.1). They, notwithstanding, at least when information control is exerted, concur to produce certain effects when they are combined with dynamic causes at a lower level. Since top-down causes, in this context, also involve dynamic causation at the lower level, they *do* have causal effectiveness. However, since they do this *through* the specific mode of dynamic causation of the lower levels, and since dynamical causes are here understood as respecting physical conservation principles (this holds true both for efficient and circular causation), there is no violation of the closeness of physical causation. Indeed, causal closure, from our point of view, is not broken if and only if principles of conservation of the relevant physical magnitudes are not violated. To avoid such a violation is sufficient to deny that the non-dynamic causes are responsible for the exchange of an additional amount of some conserved quantity. In conclusion, a careful application of the principle of closure of the physical world to different levels of complexity should lead to the result that it is impossible to *dynamically* act from one ontological level to another, either from above or from below, nevertheless, action according to goals is actually in play already at the most basic level considered here (that of molecular biology).

Top-down causes can therefore be considered as a combination of formal causes from above, material causes from below, and operations embedded in circular causes (feedback circuits) at the middle ontological level [see Fig. 1b]. In a biological context, they can very often be understood as *teleological* causation, since goals play a decisive role as far as information control is involved. In fact, the equivalence classes coming into play in information control are precisely characterized by the goals to be attained (see Sect.4.2).

**4. Functional Equivalence Classes**

**4.1 Operations and Their Conditions**
*Operations* are the elements of the functional equivalence classes we consider, here biochemical operations. Yet, an operation cannot be reduced to the pure (low-level) physical-chemical interactions *per se* but is rather a space-time pathway of such interactions; different pathways therefore define different operations. (Interactions are physical-chemical processes involving exchanges of physical magnitudes like charge, mass, energy, etc., or larger processes in which entire particles are shared or exchanged)

There are three conditions for having such operations in biology:

(1) *A space of alternative possibilities* from below (the material causes), without which one cannot have a set of different functionally equivalent operations. This set of multiple possibilities exists for micro-biological processes, since biopolymers are sufficiently complex to enter into–and be integrated in–very different forms of biological processes. For instance, catalysis can be performed by either RNA or protein molecules, two very different types of biopolymers.

*(2)* *Information selection*: In order to have a specific operation we have to select some elements from the possibility space. This is actually a selection of information, since the elements involved are biomolecules whose primary structure indeed encodes information [Lehn 2004]. Information selection is a complex modality of information that is already present at the physical level, where it occurs at the most elementary level when there is interplay between at least two quantum-mechanical systems open to an environment [Auletta 2006]. The elements indispensable to information selection are the following. (a) *Mutual information*: When several systems share a common pool of information. For instance, when there is a coupling of the receiver with the input-source. (b) *A source of variety in the input information*. Such a source of variety is very often due to random events, like many mutations in DNA, or even during epigeny or aging, and therefore plays an important role in biology. Even in cases where such variety is not random, what is important is that it is out of the control of the system it enters. This variety may be due to quantum uncertainty, or just to statistical fluctuations based in the fact that biological systems are made of many billions of discrete low-level units. The great American philosopher Charles S. Peirce [CP 1.159, 1.174, 1.302, 1.405, 5.119, 6.30-32, 6.57-59, 6.64] emphasized that in the world there is not only uniformity, but the principle of variety is "the most obtrusive character of the universe", which "no mechanism can account for". (c) *A choice (decision) event* selecting one of the incoming information variants. This might take place in a random way as in quantum measurement, where a component of the initial superposition state (the source of variety) is chosen randomly. In the biological domain, of all the environmental data (the source of variety) only certain are selected according to some criterion of relevance. In the case of microbiology, this is often through recognition of some bio-molecules by receptors and rejection of others, for example by neurotransmitter receptors in axons. This choice, however, being made between the options provided by the variety of incoming information, conveys some knowledge about the source system.

Summarizing, in all its generality we define information selection as *a decision event that selects particular input information through coupling with it*. As we shall see below, we have a second-degree of information selection, and therefore a true top-down process when a controller makes a guess and selects relative to a certain goal. Here we also need a second-level information theory, semiotics (Sec.5.3).

*(3) **Modularity***: This means there are units that are somehow *uncoupled from the external world* in that they allow information hiding, encapsulation, hierarchy, and abstraction [Booch 1994]. As internal variables are hidden, external inputs do not always determine a unique reaction of a module, for their effects depend also on its internal state. Modularity, therefore, represents a sort of divergence in the effects. The internal operations of each module are then effectively decoupled from any other same-level module. This allows that an operation can be informationally controlled *without* being dynamically causally affected by other modules or foreign factors. Modules are connected in networks which are the operative contexts where top-down causation properly takes place. It may happen that a certain network can be a sub-network (module) of a more complex network (see also Fig.3A).

**4.2 Networks and Operations**
Although the notion of network is used in a huge range of disciplines in the biological domain (see e.g. Barabas and Oltavi [2004], Schmid and MacMahon [2007]), it is usefully applicable to biochemistry only from the supra-molecular level upwards [Lehn 2004], and fits very well with our examination of equivalence classes and information control. Here, we employ a general notion of network understood as a cluster of interdependencies among units called nodes, without introducing specific assumptions about the nature either of the relations, or of the nodes.

During the constitution of networks within biological systems, information selection indeed occurs: many different molecules (the space of possibilities representing the input information) come to be coupled, becoming in this way nodes of a net in which "choice" events are initially produced, giving rise to more significant nodes (hubs) that may be considered as selecting (and gathering) information from other nodes. When this network becomes modularized, it may instantiate a specific operation.

For example, functional bio-molecular compounds such as RNA and proteins are structured entities showing a high degree of internal interconnection (see Fig.2). One can see RNA assembly and folding as a process of hierarchical and stepwise formation of a final network (the tertiary structure) through information selection, requiring first formation of specific helical elements through Watson-Crick base pairing. This defines the RNA secondary structure that is formally a network of hydrogen bonds occurring between complementary nucleotide residues (adenine (A) goes with uracil (U) and cytosine (C) goes with guanine (G)). In a second step, the presence of salts in the medium triggers the collapse of the secondary structure into structural intermediates through partial neutralization of the negative ribose phosphate backbone. Lastly, tertiary motifs that specify for particular tertiary contacts lead to the final native fold that is able to perform specific operations and therefore can carry inter-molecular recognition, catalytic, or mechanical functions. As shown in Fig.2, the final three-dimensional fold of an RNA molecule can be seen as a complex network of non-covalent interactions occurring between distant sites (or nodes) within the linear polymer sequence [Lescoute & Westhof, 2006].

The folding of biomolecules can thus be formalized as a network in which several subunits can be individuated, each of them playing a definite role within that wider context. The various subunits are determined through information selection (which picks up elements from the space of possibilities) and, spontaneously interacting, constitute a concrete complex of interrelations. Hence, the subunits are the nodes of the network and the array of interdependences among nodes specifies a determinate pathway of physical-chemical interactions, that is, a definite operation (see for example Fig. 3A). Finally, thanks to modularity, such an operation (represented by the network) is shielded against external perturbations, and then controlled and reiterated, in a top-down fashion, by a higher-level network.

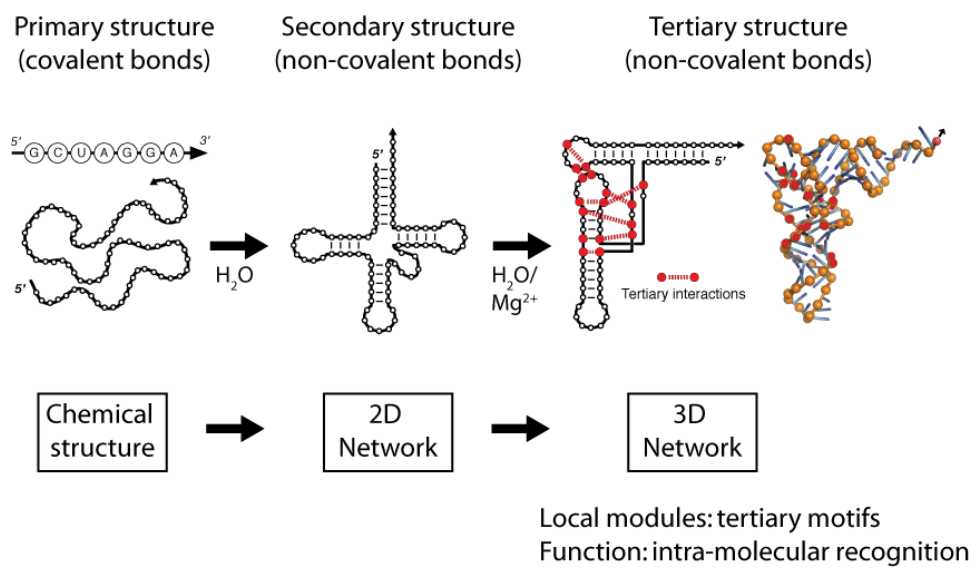

**Figure 2: The hierarchical organization of the structure of an RNA biopolymer.**
*From linear sequence information to 2D and 3D networks based on specific intra-molecular recognition. RNA biopolymers are informational and functional molecules that have the*

*ability to fold and assemble into highly intricate three-dimensional architectures that take advantage of networks of specific tertiary interactions [Lescoute & Westhof, 2006].*

---

Another example of a network is provided by interactome pathway maps, and as an isolated operation we may consider endocytosis, the operation through which some external molecule will be engulfed in a cell [Schmid and McMahon, 2007].

In many cases, some of the nodes of a network can be substituted by different suitable sets of chemicals without losing the overall features of the network itself. It is noteworthy that some nodes can be simply dropped without altering the effect of the whole pathway. This shows that the same function can be instantiated through different clusters of physical-chemical interactions.

**4.3 Equivalence Class Definition**
Mathematically, an equivalence relation is a type of relation on a set that provides a way for elements of that set to be identified with other elements of the set. Those elements considered equivalent through this identification form an equivalence class.[3]
Let $W$ be a set and let $w$, $x$, and $y$ be elements of $W$. An *equivalence relation*, ~, on $W$ is a relation on $W$ that is,
  **Reflexive**: $w$ is equivalent to $w$ for all w in $W$.
  **Symmetric**: if $w$ is equivalent to $x$, then $x$ is equivalent to $w$.
  **Transitive**: if $w$ is equivalent to $x$ and $x$ is equivalent to $y$, then $w$ is equivalent to $y$.

If $W$ is a set, $w$ an element of $W$, and ~ an equivalence relation on $W$, the *equivalence class of w* is the set of all elements of $W$ equivalent to $w$ under ~.

**4.4 Functional Equivalence Classes**
As we have said, the concept of function is now widely acknowledged in biology. However, we emphasize that a function defines an equivalence class. When one speaks of equivalence in general, one understands a "possibility of substitution in any context". It is fundamental that ***functional* equivalence classes** are on the contrary context-sensitive. The concept of equivalence class that is the object of this paper is therefore not a formal-logical construct but a pure functional-biological category, where different operations are considered functionally equivalent if they produce the same outcome for some functional purpose (the goal). Thus, we focus on *functional* equivalence classes rather than purely formal equivalence classes, even if the properties defined previously must also hold for them.

To be specific, let us consider the biochemical subsystems that are required to synthesize a minimal cell: genome replication, transcription, translation, correct protein folding, and any necessary post-translational amino-acid modifications [Forster and Church, 2007]. Some of these abstract functions can be fulfilled by different pathways of physical-chemical interactions, i.e. by different operations (see Fig.3). When different operations fulfill the same function, they form an equivalence class (in this case the formal relations of Section 3.3 will be fulfilled). Since operations equivalent with respect to a certain function *are not* automatically equivalent for other functions, it is important to identify unequivocally the function concerned. Therefore, the criterion by which items are judged to be or not to be members of such classes is *only* a specific function. In this way, functional equivalence classes are characterized by a *part-whole* relation, which, in turn, links a particular kind of function directly with the issue of signs (see Sec.5.3 below). It should be noted that good examples of biological systems belonging to a particular functional equivalence class are evolutionarily unrelated functional bio-molecules that result from convergent evolutionary processes. A clear example of functional equivalence class is provided by splicing, which is the process involved in the transition from pre-mRNA to mature mRNA. The function "splicing" is performed through different operations (making use of different chemicals, see Fig.4). Despite such a difference, the outcome is the same as far as the splicing is concerned.

---

[3] Equivalence classes (see http://www.iscid.org/encyclopedia/Equivalence_Class) are based on equivalence relations (see http://www.iscid.org/encyclopedia/Equivalence_Relation).

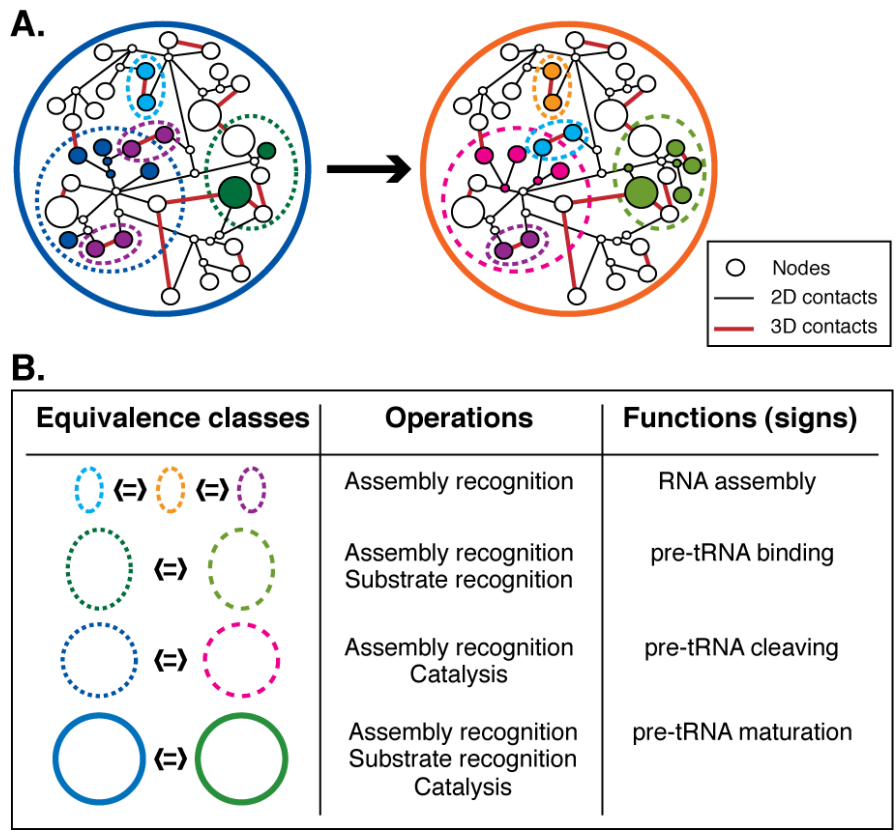

**Figure 3: Equivalence classes, operations, and functions.**
**3(A)** *Two functionally equivalent networks. Different nodes and contacts specify for different operations. Modularity is readily apparent as some local nodes and related contacts are found at multiple locations within each network. The whole network can be seen as a hierarchical organization of smaller interchangeable lower order networks.*
**3(B)** *Examples of operations and functions (see also Sec.2). If the two networks in A are taken as the three-dimensional structure of two RNA biopolymers that belong to the same functional class (e.g. Rnase P RNA), then some of their constitutive structural modules belong to lower-level functional equivalence classes.*

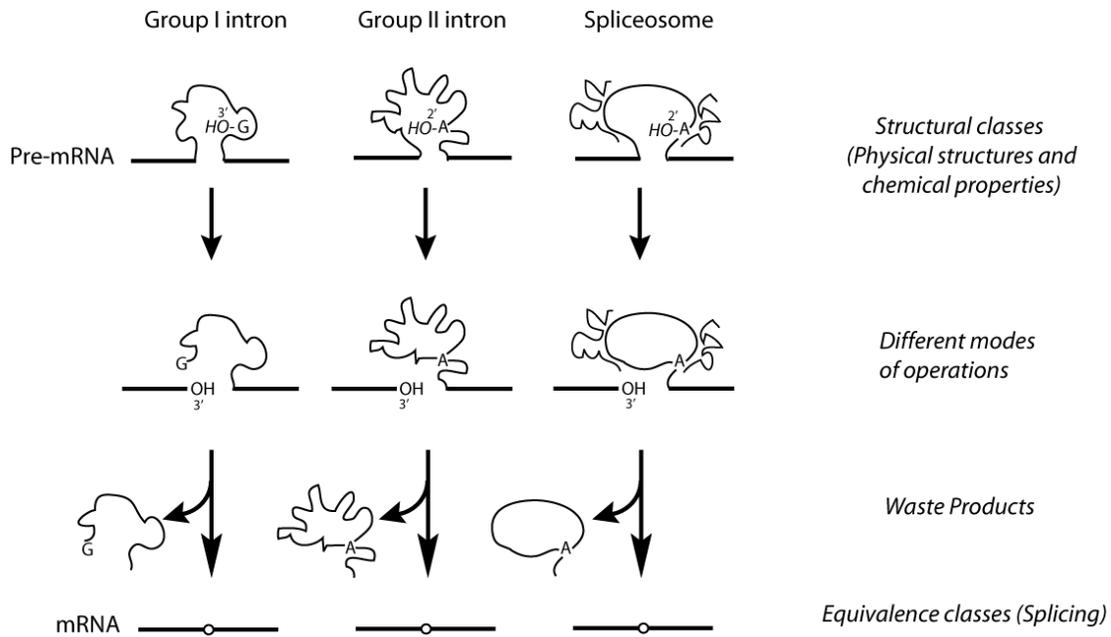

**Figure 4: Splicing.**
*Splicing as an example of equivalence classes. Here, the fact that the different modes of operations indeed provide mRNA (as an outcome, and independently of the operations that mRNA can, in turn, perform relative to a further function) points out that these modes pertain to an equivalence class.*

---

Functional equivalence classes can contain multiple modules belonging to functional equivalence classes of lower complexity (Fig.3). The demonstration that lower functional equivalence classes are under hierarchical control of higher-level ones can be seen as evidence of top-down causation. We recall that in order that a higher-level network can be considered as constituted by lower level sub-networks, each of them must produce the outcomes required by the functionality of the higher network, no matter what modes of operation they use to produce it. The way in which this hierarchical control is exerted is by means of information feedback control as we will see in the next section.

During the constitution of an equivalence class, and in general when top-down causation through information control is at work, one can expect selection of fewer different types of structural motifs belonging to an equivalence class of larger networks than one would normally expect given the mere combination of lower-level elements (even smaller networks). This is due to the fact that the information of the network of higher complexity strongly affects the nature of information of the lower nodes, reducing the number of possibilities as the number of constraints within the larger system increases. The best analogy we have for RNA is that of a language. With RNA, it is possible to speak many "words" that have the same "meaning", as they belong to an equivalence class. When these words are linked to one another, they define a "sentence" that conveys a greater meaning than its more basic components. Now the question might be whether it is the information of the words at the lower level that is affecting the information carried by the whole sentence, or the opposite. In fact, one can realize that what may be critical for the information carried by the whole sentence is the syntax that connects the words together, rather than the words themselves.

## 5. Information Control

### 5.1 What is Information Control?
Since we are focusing on top-down causation by information control, we have now to see how information control should be characterised.

One of the biggest misunderstandings in information theory is to have taken Shannon's [1948] theory of communication (in the context of controlled transmission) as a general theory of information. In such a theory, centred on signal/noise discrimination, the message is already selected and well defined from the start, since from the start the selection operation among several alternative states (or bits) has already occurred (at the input or sender), and the problem here is only to faithfully transmit or further process, in the presence of disturbances, the sequence of bits that has been selected [Auletta 2007b]. On the contrary, a true information theory (as was Wiener's [1948] original aim) starts with an input as a *source of variety* and has the selection only *at the end* of the process. In other words, a message here is only the message selected by the *receiver*. As a matter of fact, *any* information reception will be subject to the original variety, in addition to the consequences of disturbance, dispersion, or even of practical needs, and use of any of this information, at the most elementary level, already constitutes information selection. This is momentous for biological systems, since they are confronted with an environment that represents sources of uncertainty, and for this reason do not have control from the start of the string of bits that has been sent. Even inside a single cell we have such a problem, due to the modularisation of the different subsystems. The control process must here somehow be constructed while having only a limited pool of resources.

Let us state the problem in this way. According to the traditional information (communication) theory, the main problem is reliability, understood as the matching between input and output. However, in biological processes we are much more interested in situations in which the receiver does not have full control over the input and is therefore *forced to guess the nature of the input by taking the received partial information as a sign of it* (revealing its nature). This provides the more basic condition for equivalence classes, in that it is possible to take a partial input as a sign of many (possible) different entire input situations, and to regard different inputs as equivalent under a certain point of view.

At any biological level, the receiver is in general flooded with incoming data, and has to separate *background data* (important but constant) and *noise* (irrelevant data) from *relevant information* (data that are needed for some purpose). Therefore, *information control* consists in information selection–often involving a guess–from a certain point of view, and this represents the goal of the system. For instance, a bacterium searching for an energy source may use a specific temperature gradient (the received information) as a sign (see below Sec.5.3) of this source. Obviously, many different temperature distributions (within a certain window) will fit, and therefore allow it to reach a certain source, which is the goal. Moreover, any source that fits some general criterion will be good, and therefore pertains to the equivalence class established by the goal of acquiring energy. We need now to state how goals and feedback control are linked. Information control via feedback is not the only way to have control via information but it plays a fundamental role in living systems: It is indeed involved, for example, in any homeostatic process that a living system must perform to survive.

**5.2 Goals and Feedback Control**
We can speak of top-down causation by information control when there are two elements that cannot be reduced to any low-level explanation: 1. the *formal structure* determining the feedback control loop (the formal cause, in our language), and 2. the *goal*. These two elements represent the way the equivalence class is built in and controlled from above. To exercise information control, we need to join a system (the controller) with another system (the performer), see Fig.5. However, not all forms of connection will work, but only those satisfying the following requirements:
   (a) the performer has to execute an operation in order to deploy the function needed by the controller, and
   (b) the controller needs to be able to verify step by step if the function is actually deployed to the required degree.

The requirement (a) is strictly linked to the fact that the controller has an inbuilt goal to reach; the requirement (b) is fulfilled by a feedback circuit (and an inbuilt comparator). It seems that mechanical devices, like a thermostat, are able to implement information control without any intervention of biological elements and in purely mechanical terms. This is however an erroneous point of view, since such devices have been built by humans to act in a certain way. Therefore, the functional element (and the goal) is already inbuilt.

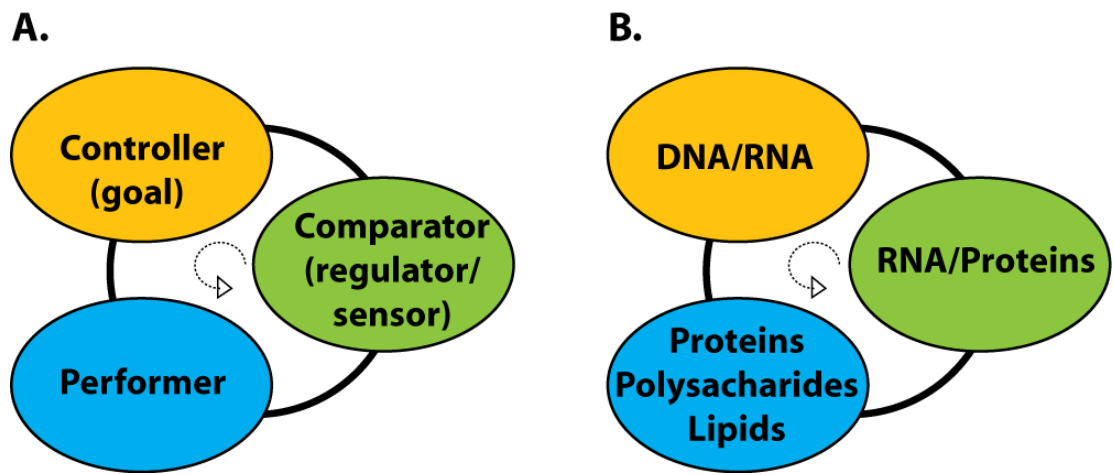

**Figure 5: The basic feedback control process.**
*5.A: In living beings, feedback control underlies homeostatic processes (maintaining blood pressure, body temperature, etc.). The comparator determines the difference between the system state and the goal, and sends an error signal, activating the controller to correct the error. This is the way that goals (abstract variables) control dynamic causes, thus becoming causally effective. The goals in simple biological systems are genetically determined (unchanging through the organism's life and constant across a species).*
*5.B: The application of this framework to present-day cellular systems. We show the biopolymers involved in each step.*

*Feedback control* (see again Fig.5) is therefore the way information is causally effective in the biological world [Milsum 1966, Calow 1976, Ellis 2006a]. Feedback structuring is necessary in making information useful for attaining goals, otherwise the controller could not recognize whether it had succeeded in obtaining a desired outcome. In this way the controller is able to use a specific operation by a performing system as an element of a functional equivalence class.

At lower biological levels, the goal is somehow built into the controller and does not itself need to be "chosen" (the situation is different when consciousness is involved). The goal-state (relative to a certain specific parameter) is not a need *for the controller itself*, an entity which is a part, a *sub*-system, of the whole biological system. Moreover, that the goal-state represents a need for the whole biological system does not mean this is an end *in itself*, since in general the exigency of keeping the system-state coincident with the goal-state depends on the latter being crucial for *other* processes and functions, in turn indispensable for the whole system. An example of information control is the expression or the repression of segments of DNA. Indeed, this control procedure, done on strings containing information and through activators and repressors that carry and instantiate instructions, is highly contextual and depends on the goals that the organism pursues in different time windows and states of its developmental or metabolic activity.

When an operation is actually performed, the fact that the outcome satisfies the goal (so that the comparator no longer sends error signals), can be seen as evidence for the operation's pertaining to the equivalence class defined by the goal itself (this is the reason why functional equivalence class is not only an epistemic category). The goal can be satisfied by several slightly different outcomes, all of them falling into a well defined range of tolerance. In order for an operation to be acknowledged as a member of a functional equivalence class, it is sufficient to individuate a specific feature that is tightly bound with the outcome achievement. Therefore, when we speak of equivalence classes at a biological level (*functional* equivalence classes), we enter the domain of signs (see Sec.5.3). That the goal defines the equivalence class related to it is reminiscent of the concept of equifinality, a type of convergence, introduced by von Bertalanffy [1969] in the framework of general system theory.

**5.3 Signs and Equivalence Classes**
The connection between information control (and top-down causation) and equivalence classes, is exactly represented by the goals and the related issue of signs.

In our usage, a **sign** is something that stands for something else in a certain context, or for a certain goal [Peirce CP: 2.228 and 1.540; Auletta 2007a, 2007c], and this is what establishes equivalence classes in a context of information control. Signs convey information that has to be recognised, and this occurs even in biological systems without consciousness [Hoffmeyer 1996; Deacon 1997].

As we have seen, in any information exchange we have selection *at the end*, not at the start. Signs, in the most elementary case, take this output selection as saying something *about the input*, so that the receiver starts a new information process aiming at the source (inverting somehow the ordinary flow of information from the source to the receiver, Fig.6). There are many examples of the use of signs in biology. Amongst the most remarkable ones are the so-called *affordances* [Gibson 1966,

1979], when an animal takes a physical input (a smell) as sign of something that is fundamental for survival (food).

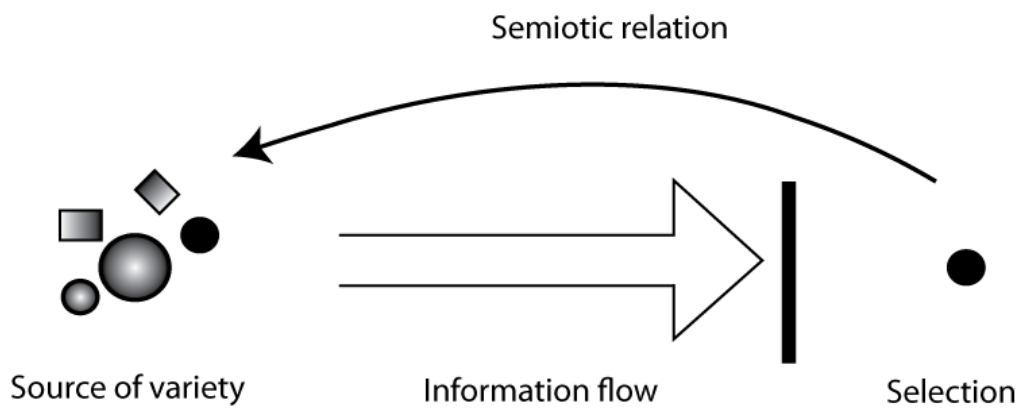

**Figure 6: Information and signs.**
*Input information, as a source of variety, starts an informational process that is concluded when information selection is accomplished. When this selected element is taken to say something about the input information, we have a semiotic relation and say that it is a sign of the input. This is evidently true when this element is tightly connected with that complex of initial information; something, however, can also be a sign of things that are not obvious consequences of the input information, for instance when certain items are taken to be a sign of the needed element (able to satisfy the goal).*

With respect to equivalence classes, the *outcome* of the operation performed in a certain biological context may be considered as a sign of the function being deployed successfully (or not), and hence that the operation really is (or not) a member of the equivalence class. In other words, of all the input information entrained in the physical-chemical properties of the molecule or of the interaction under consideration, a *single* feature is selected and taken as sign of the function required by the inbuilt goals of the whole system (as we have said, it is a part-whole relation). This sign may be the outcome of the operation, or any of its features reliably associated to the outcome. Let us call this sign a *mark* of the operation. Systems able to perform a certain operation for a certain goal are acknowledged through a specific mark, and controlled in their mode of operation through such a mark. Thus, the informational control of an operation (see Sec.5.2) is a genuine semiotic process, since the control instance needs to catch and select specific information as a sign of the fact that things are going in the right or the wrong way.

**5.4 Functional Selection and Top-Down Causation**
In many cases, new information is acquired through functional selection–and therefore specialization–of determined operations at the lower level. Strong evidence for this process is given by mutation and heritage of epigenetic mechanisms, like the restoration of native folding of single stranded DNA sequences through reverse mutations [Shepherd et al, 2006]. When such a specialization occurs, it becomes truly useful for attaining a certain higher-level goal, once it is linked to the previous set of controllable information of the biological system through a feedback loop, eventually establishing and maintaining a new functionality. Such a process is also known, at evolutionary scale, as *exaptation* and consists in general in the use of adaptive traits in a new context that is different from that in which they have originally arisen [Gould and Verba, 1982]. This situation also enables *biological systems to undergo adaptive selection at a functional level*.

An excellent example is the transition from an RNA-based genome to the DNA-based genome [Jaeger & Westhof, 2001; Lehninger et al, 2004]. RNA shows two fundamental chemical instabilities: (1) Instability of the ribose-phosphate backbone that can hydrolyse inherent to the presence of a hydroxyl group at the level of the sugar moiety, and (2) instability of cytosine that can hydrolyse into uracil. At this level, these instabilities prevent the genome from growing too much due to an increasing loss of information through these chemical instabilities, and this represents a limitation of information coding. To circumvent this major problem, two new metabolic pathways have been developed through adaptive processes. One corresponds to the formation of deoxy-nucleotides from nucleotides, removing the first instability. The second one is much more remarkable, and consists in the transformation of uracils into thymines by addition of one carbon methyl to the base. The last change allows a mechanism to repair a damaged or mutated genome to emerge.

This key step in evolution provides a better vehicle of information coding (DNA) as well as allowing RNA to be more dedicated to operative tasks rather than to long-term storage functions within the cell. It also provides the cell with a better control over the expression of its various functional components (DNA, proteins, and RNA) by allowing modulation and better timing control. In other words, this increase in information control corresponds to an increase of the modularity of the biological system by specialization of its functional tasks. Obviously, such a *specialization* may work only if it happens in a suitable network of information control.

**5.5 Our Whole Conceptual Framework**
Top-down causation by information control occurs thanks to the connection between equivalence classes and information control. We have *top-down causation* by information control when, once an equivalence class has been established, the information selection defining the operations which are the elements of the class is conserved, thanks to modularity, despite the variability of lower level variables. In this case, the feedback control circuits produce reliable responses to higher-level information [Ellis 2006b, 2008], allowing equivalence classes of lower level operations that give the same higher-level response for a certain goal. As equivalence classes are abstract configurations rather than physical states, and are (through the lower level modality of operations) the causally effective higher-level entities in this context, we have top-down causation that cannot be reduced to any specific bottom-up or same-level causation. The system as a whole is illustrated in Fig.7.

**6. Biological Modelling**

**6.1 A Model of an Equivalence Class**
The concept of equivalence class is readily available at the level of stable RNA biopolymers. Sequence and structural analysis of natural RNA molecules such as large self-folding catalytic RNAs (group I and group II introns, RNase P RNA, ribosomal RNAs) reveal that their conserved structural catalytic cores are often stabilized by peripheral modules or tertiary motifs that, despite their different local structures, contribute in a similar fashion to the final core assembly and stabilization [Jaeger *et al.* 1994, Jaeger *et al.* 1996; Westhof *et al.* 1996, Massire *et al.* 1997; Massire *et al.* 1998].

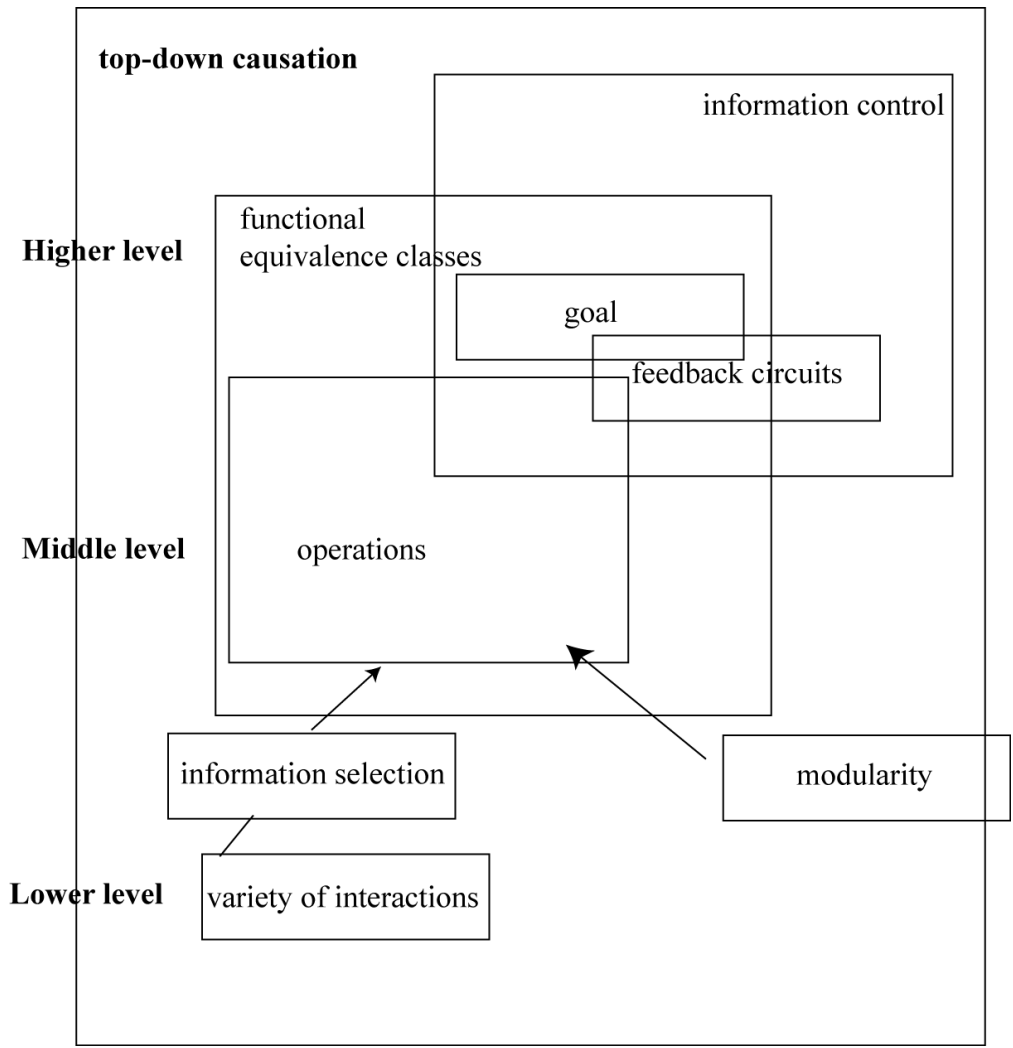

**Figure 7: Top-down causation.**
*The relations between the elements involved in our theoretical scheme. The arrows show the conditions that contribute to the establishment of an operation.*

As shown in Fig.8.A, the 2D structural networks of two different molecules of RNase P RNA have a conserved structural core able to perform the same catalytic function that is the maturation of tRNAs (see also Figs.2-3). They however differ in their peripheral regions, which are involved in the proper intra-molecular recognition of their two constitutive domains. As seen in Figs.8.B and 8.C, this intra-molecular recognition function that allows spatially forming a tertiary contact between the helical element P9 and the helical element P1 is operated by two kinds of tertiary structural motifs that have different sequences and three-dimensional structures. For this reason, the modality of operations is also different. However, these two specific operations (and the related different tertiary motifs) are perfect examples of items belonging to the same equivalence class (RNA-RNA recognition). These components have been shown to be swappable within the structural context of a group I ribozyme, demonstrating that they are functionally equivalent [Jaeger *et al.* 1994].

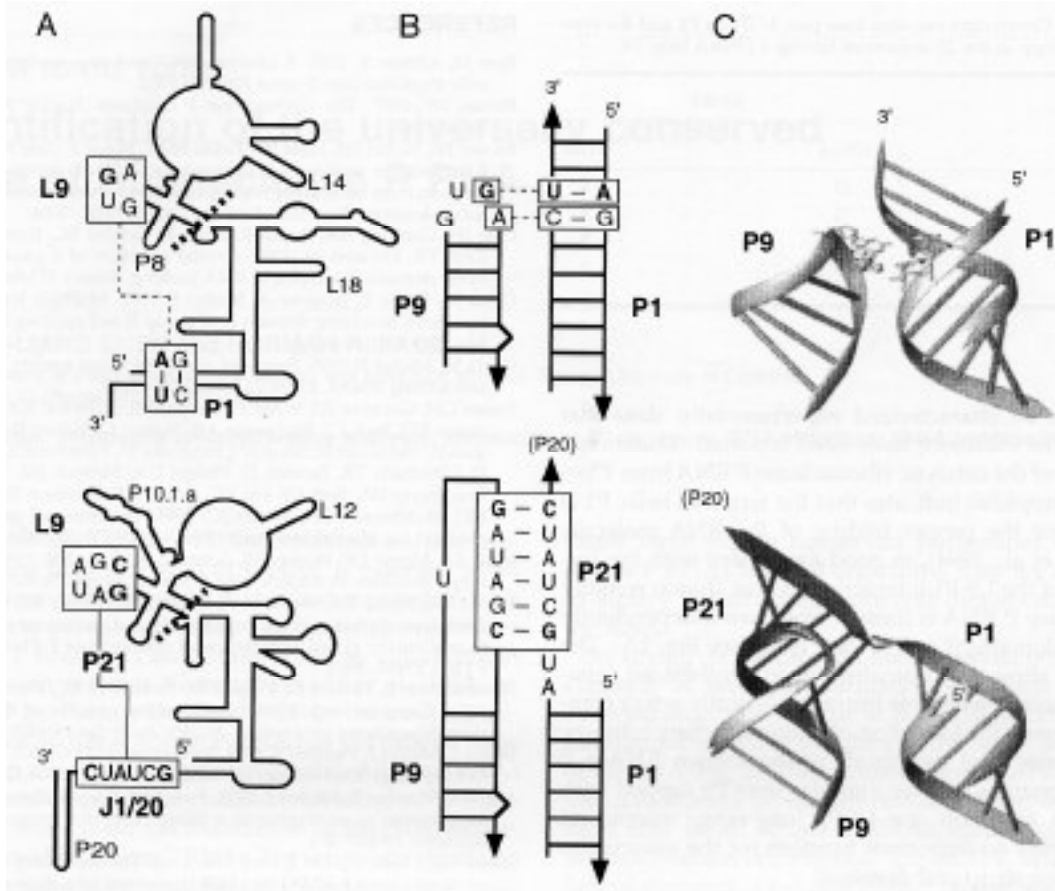

**Figure 8: Operations and Functional Equivalence.**
*Different operations connected with differences in networks that are notwithstanding functionally equivalent (Adapted with permission from Massire et al, 1997). Top: RNase P RNA of type A, Bottom: RNase P RNA of type B.*

**6.2 Experiments on Top-Down Causation**
Notwithstanding the interesting results reported in Sec. 6.1, we wish to find stronger evidence for top-down causation by showing the establishment of equivalence classes by systems *displaying information control*. In Sec.2 we have reported about an important experiment showing that, notwithstanding differences at the physical-chemical level, enzymes pertaining to different organisms can be interchanged without losing their functionality. In our opinion, their experiment is a beautiful instance of the fact that different lower-level operations are *informationally controlled from above* and constitute or may constitute functional equivalence classes.

For this reason, the experiment by Wegscheid et al. can be framed within a general model for testing top-down-causation. Here (see Fig. 9.A), one takes a unicellular organism *X* and identifies an operation *a* (we recall that an operation is a pathway of physical-chemical interactions) deploying a certain function in *X* (for example, the maturation of tRNA). Then, one takes another organism *Y* deploying an alternative operation *b* and substitute it for *a* in *X*. Then, one considers whether the operation *b* is able to deploy the same function in the new context. If operation *b* is actually "accepted" by the organism *X*, then the role of functional equivalence class for that organism is proved to the extent to which the specific interaction-pathway is disregarded and only the function taken into account.

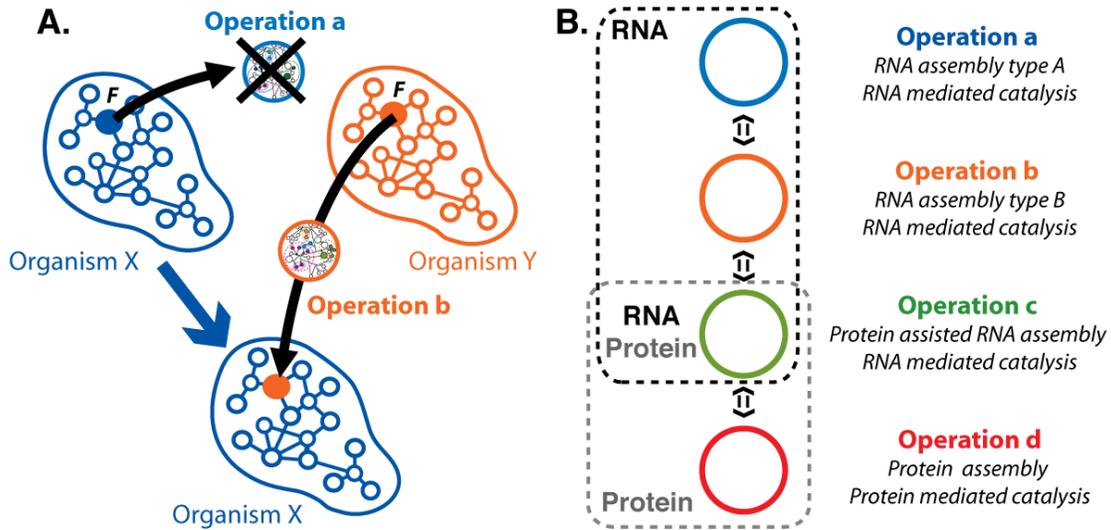

**Figure 9**: **Demonstration of functional equivalence within the metabolic network of an unicellular organism**. *Colours here do not follow the previous conventions.*
**9.A**: *See text.*
**9.B**: *The same function can be carried out by molecules with different operations. Operations a and b, which are carried out by RNA alone, are different due to the way they fold. Operation c takes advantage of a protein to help the RNA fold. Operation d is mediated by protein alone at the assembly and catalytic levels.*

This means that the equivalence class in this case is established inside a *single* organism, suggesting that this happens through an information-control feedback loop. Notice that trans-individual equivalence classes (as in the experiments reported in Sec. 6.1) could not be considered as evidence for top-down causation but rather as a way the scientist classifies certain physical and chemical interactions. In fact, in this case what happens within two (or more) different organisms could be explained in purely mechanical and bottom-up terms in each organism separately. When, on the contrary, a single organism *X* is able to deploy a certain function either by performing operation *a* or *b*, that can be no longer understood in a purely mechanistic way, and what becomes of primary importance is the goal the operation is aiming at (and the control instance piloting the process).

However, within the context of the Wegscheid experiment, it is important to stress that while type A and type B RNase P have significantly different architectures, they are still evolutionary related as their secondary structure and tertiary structure share a consensual catalytic site for binding and cleaving tRNA molecules. The two architectures are both resulting from the evolutionary divergence of a common RNase

P RNA ancestor. This is clearly established by extensive sequence comparative analysis and crystallographic analysis of the structure of the two classes of RNase P RNA [e.g. Altman 2007, Kirsebom 2007]. Type A and type B RNase P enzymes have evolved through time to retaining in their common structural core all the specific structural determinants for performing the same catalytic reaction of maturation of pretRNAs. Therefore, a reductionist explanation such as the one proposed by Wegscheid et al for their experiment cannot be completely ruled out.
The basic scheme of the type of experiment proposed is quite common in genetics, when a gene coding for a certain function is substituted by another similar one presenting some genetic variations.

Therefore, what we really want to perform is an experiment of substitution of molecules of the same function but with different modes of structural and functional operations, meaning different ways to perform the catalysis and recognize the substrate. This would show that two different evolutionarily unrelated molecular systems that share the same function by convergent evolution can substitute for one another. In fact, in living cells, there are numerous examples of such types of systems (e.g. DNA nucleases, amino acyl tRNA synthetases, Self-splicing introns, self-cleaving ribozymes…).

Therefore, a further refinement of the experiment (Fig. 9.B) described above could be to eventually substitute a key cellular operation performed by a protein (like a nuclease activity) by one performed by an RNA. The functional RNA does not have to be of natural origin as it can eventually result from *in vitro* experiments [e.g. Jaeger 1997, Chworos & Jaeger 2007]. In this case, while the function could be the same (this is our prediction), any level of organization below that one would be different, allowing demonstration of top-down causation by information control without ambiguity. Here, we assumed that the organism-level is higher than that of operations, and therefore anything that could be swapped, maintaining the same cell activity and its outcomes, is seen as an effect of top-down causation.

**6.3 A Research Program**
With this paper we propose a new research program, through which we try to experimentally establish top-down causation. It is obviously a very difficult issue, and probably many experimental steps as well as new types of experiments will be necessary. However, we also believe that, given the conceptual framework developed here, it is a possible enterprise.

It is always difficult to positively prove a result, so one may wonder whether it is possible to positively prove top-down causation. In complex cases, one could always object that there will be at some time a purely molecular (and bottom-up) explanation of the eventual findings we are aiming at. However, showing that equivalence classes are constituted at the most basic bio-molecular level represents a strong counterexample to all bottom-up and same-level explanations, strongly suggesting top-down causation as a fruitful way to understand these results. It will also help to explain issues regarding emergence and related evolutionary aspects that constitute the major part of our future programme. It should be also clear that a possible failure of our research does not necessarily imply a direct confutation of top-down causation. Other ways to search for it, both at the microbiological level and at higher levels of complexity, are likely to be found.

**Acknowledgement**. We wish to express our warmest gratitude to Ivan Colagé and Paolo D'Ambrosio, two doctoral students of the Pontifical Gregorian University, who actively participated in the discussion developed during the workshop held in Rome on September 24th-28th that made this paper possible. We thank Horst Klump, of the University of Cape Town, for the example of native folding of DNA sequences, and Bill Stoeger, of the Specola Vaticana, for helpful comments. The paper has been significantly improved through constructive comments of our referees.

Luc Jaeger wishes to dedicate this paper to the memory of Father Benvenute Bavaro, OFM.